\documentclass[twocolum,aps,prc,reprint,superscriptaddress,floatfix,article]{revtex4-2}

\usepackage{graphicx}
\usepackage{subcaption} 
\usepackage{amsmath} 
\usepackage{amssymb}
\usepackage{fleqn}
\usepackage{titlesec}
\usepackage{color}
\usepackage{dcolumn}
\usepackage{bm}
\usepackage{mathtools}

\begin{document}

\preprint{APS/123-QED}

\title{\textbf{Effect of neutron-proton asymmetry on the $^3$H clustering in Boron isotopes} 
}%

\author{J. L. Jin}
\affiliation{%
 College of Physics and Technology, Guangxi Normal University, Guilin 541004, Guangxi,  China
}%
\affiliation{%
 School of Science, Huzhou Normal University, Huzhou 313000, Zhejiang, China
}%

\author{Q. Zhao}%
 \email{Contact author: zhaoqing91@zjhu.edu.cn}
\affiliation{%
 School of Science, Huzhou Normal University, Huzhou 313000, Zhejiang, China
}%

\author{P. J. Li}
\affiliation{%
 Key Laboratory of Nuclear Physics and Ion-beam Application (MOE), Institute of Modern Physics, Fudan University, Shanghai 200433, China
}%
\affiliation{%
 Shanghai Research Center for Theoretical Nuclear Physics, NSFC and Fudan University, Shanghai 200438, China
}%

\author{M. Kimura}
\affiliation{%
 RIKEN Nishina Center for Accelerator-based Science, RIKEN, Wako 351-0198, Japan
}%

\author{D. Beaumel}
\affiliation{%
 Université Paris-Saclay, CNRS/IN2P3, IJCLab, 91405 Orsay, France.
}%

\author{B. Zhou}
\affiliation{%
 Key Laboratory of Nuclear Physics and Ion-beam Application (MOE), Institute of Modern Physics, Fudan University, Shanghai 200433, China
}%
\affiliation{%
 Shanghai Research Center for Theoretical Nuclear Physics, NSFC and Fudan University, Shanghai 200438, China
}%

\author{J. L. Tian}
 \email{tianjl@gxnu.edu.cn} 
\affiliation{%
College of Physics and Technology, Guangxi Normal University, Guilin 541004, Guangxi,  China
}%
\begin{abstract}
To investigate the influence of neutron-proton asymmetry on the formation of asymmetric clusters, we perform a systematic comparative study of $^{3}$H and $\alpha$ cluster preformation in the Boron isotopic chain ($^{11-14}$B). Within the framework of Antisymmetrized Molecular Dynamics (AMD), we compute the nuclear wave functions and subsequently extract the reduced width amplitudes (RWA) and spectroscopic factors (SF). The results show that the $\alpha$ cluster SF exhibits a monotonic decrease with increasing neutron number, consistent with the established suppression effect of the neutron skin. In contrast, the $^{3}$H cluster SF displays a non-monotonic behavior, peaking at $^{12}$B. This distinct trend indicates that the formation of the asymmetric $^{3}$H cluster is subject to a competition between suppression from the neutron skin and an enhancement driven by the neutron-proton asymmetry of the parent nucleus. We successfully isolate this enhancement effect by analyzing the ratio of the SFs, SF($^{3}$H)/SF($\alpha$). This approach not only quantifies the enhancement but also proposes the SF ratio as a robust experimental observable for probing insights into asymmetric clustering phenomena.

\end{abstract}

\maketitle

\section{Introduction}
Nuclear clustering is a fundamental structural phenomenon in light nuclei, which provides crucial insights into the interplay between the nuclear force and many-body correlations. Traditionally, the $\alpha$ cluster, characterized by its remarkable binding energy and spin-isospin symmetry, dominates cluster configurations in stable or near-stable nuclei~\cite{Ikeda1968, Freer2018, Horiuchi1991}. The formation and decay of $\alpha$ clusters play pivotal roles in understanding nuclear stability, shell evolution, and astrophysical reaction rates~\cite{Oertzen2006, Tanihata1985}. More recently, it has become widely recognized from both theoretical and experimental perspectives that the presence of a thick neutron skin dilutes the spatial density at the nuclear surface~\cite{Tanihata1992Revelation, Brown2000Neutron}. This effect hinders multi-nucleon localization, which in turn leads to a systematic reduction in the preformation probability of symmetric $\alpha$ clusters~\cite{Yoshida2018Investigation, Tanaka2021Formation}.

However, as the exploration of the nuclear landscape extends toward the neutron drip line, the increased neutron-proton asymmetry introduces new degrees of freedom to the traditional clustering picture. An intriguing question arises: how does the neutron-rich environment affect the formation of asymmetric clusters, such as the $^{3}$H cluster? Unlike the $\alpha$ particle, the $^{3}$H cluster intrinsically possesses a neutron-proton asymmetry. Therefore, the preformation probability of the $^{3}$H cluster in a neutron-rich nucleus is governed by a delicate competition. On one hand, similar to the $\alpha$ cluster, it is subject to the suppression effect from the neutron skin. On the other hand, the overall neutron excess may structurally favor the clustering of an asymmetric subsystem, an effect driven by the neutron-proton asymmetry of the parent nucleus. Disentangling these two competing effects is critical for understanding the molecular-like structures and exotic clustering behaviors far from the line of stability.

The Boron isotopes provide a suitable system for investigating this competition. From the perspective of the cluster model, the five protons in Boron isotopes naturally suggest a cluster configuration involving two $\alpha$ clusters and one $^{3}$H cluster ($\alpha+\alpha+^{3}$H). More importantly, while one $\alpha$ cluster can be associated with the $s$-shell, the other $\alpha$ and the $^{3}$H cluster must both form under similar $p$-shell mean-field conditions, thereby placing them in the same competitive environment. Therefore, from both cluster and shell model perspectives, the Boron isotopes provide a system of moderate complexity that is sufficiently representative for studying the differences in the formation of these two cluster types.

In this work, we employ the framework of Antisymmetrized Molecular Dynamics (AMD)~\cite{KanadaEnyo2003, KanadaEnyo2012, Kimura2016Antisymmetrized} to calculate the wave functions of the relevant nuclear systems. From these wave functions, we then compute the reduced width amplitude (RWA) and the corresponding spectroscopic factor (SF)~\cite{Chiba2017Laplace, Zhao2021Alpha}. The SF serves as a quantitative measure of the cluster preformation probability. Crucially, it is an observable that can be measured in nuclear reaction experiments, providing a direct means to test the theoretical results presented herein. Furthermore, we propose that the influence of neutron-proton asymmetry on $^{3}$H cluster formation can be effectively isolated by analyzing the ratio of the SFs for $^{3}$H and $\alpha$ clusters.

The present paper is organized as follows. In Sec.~\ref{sec:theoreticalwork}, we briefly introduce the theoretical framework, including the AMD framework, and the formulations for calculating RWA and SF. In Sec.~\ref{sec:results}, we present the calculated ground state energies, the behavior of RWAs, and the detailed discussion on the SFs and the isospin-enhancement mechanism. Finally, a summary is provided in Sec.~\ref{sec:summary}.

\section{Theoretical framework}
\label{sec:theoreticalwork}

In this paper, we adopt the microscopic many-body calculation to calculate the wave functions of the nuclei. The Hamiltonian for the calculations is given as:
\begin{equation}
H = \sum_{i}^{A} \hat{t}_i - \hat{t}_{c.m.} +  \sum_{i<j}^{A} \hat{v}_{NN}(ij) + \sum_{i<j}^{Z} \hat{v}_{C}(ij),
\end{equation}
where the nucleon-nucleon ($NN$) interaction $\hat{v}_{NN}(ij)$ is the Gogny D1S interaction~\cite{Decharge1980Hartree}. The Coulomb interaction $\hat{v}_{C}(ij)$ is approximated by a superposition of $7$ Gaussian functions~\cite{KanadaEnyo2012, Suhara2010}. The center-of-mass kinetic energy term $\hat{t}_{c.m.}$ is subtracted from the total energy.

The basis wave function is a parity-projected Slater-determinant~\cite{KanadaEnyo1995, KanadaEnyo2003}:
\begin{equation}
\Phi^\pi = \hat{P}^\pi \mathcal{A}\{\varphi_1\varphi_2\cdots\varphi_A\},
\end{equation}
where $\hat{P}^\pi$ is the parity projection operator, and $\mathcal{A}$ is the antisymmetrization operator to restore the anti-symmetry of the particles. $\varphi_i$ is the single-particle wave function, which is represented by the Gaussian as
\begin{equation}
\begin{aligned}
\varphi_i(\boldsymbol{r}) = \exp\left\{ -\sum_{\sigma=x,y,z} \nu_\sigma \left(r_\sigma - Z_{i\sigma}\right)^2 \right\} \chi_i \tau_i~,\\
\chi_i = a_i \chi_\uparrow + b_i \chi_\downarrow,\quad \tau_i = \{\text{proton or neutron}\}~,
\end{aligned}
\end{equation}
where $\boldsymbol{Z}_{i}$ is the centroid of the Gaussian. $\chi$ and $\tau$ are the spin and isospin parts of the nucleon, respectively. The parameters, $(\nu_x,\nu_y,\nu_z)$, $\boldsymbol{Z}_{i}$, and the coefficients $(a_i, b_i)$ for the spin part, are determined via the frictional cooling method~\cite{Kimura2004}, which minimizes the sum of the Hamiltonian eigen value and the constraint potential energy:
\begin{equation}
E(\beta) = \frac{\langle \Phi^\pi|H|\Phi^\pi \rangle}{\langle \Phi^\pi|\Phi^\pi \rangle} + v_\beta(\langle \beta \rangle - \beta)^2~.
\end{equation}
The strength of the constraint potential $v_\beta$ needs to be sufficiently large to ensure the deformation of the basis wave function $\langle \beta \rangle$ is equal to the input value $\beta$ by minimization of $E(\beta)$. Under this premise, the basis wave function $\Phi^{\pi(\beta)}$ will be the optimized wave function with the lowest energy corresponding to the given deformation parameter $\beta$.

After the deformation constrained, the angular momentum projection is performed to restore the spin of the nucleus, 
\begin{equation}
\Phi_{MK}^{J\pi}(\beta) = \frac{2J+1}{8\pi^2} \int \mathrm{d}\Omega D_{MK}^{J*}(\Omega) R(\Omega) \Phi^\pi(\beta)~,
\end{equation}
where $D^J_{MK}(\Omega)$ and $R(\Omega)$ denote the Wigner's D-function and the rotational operator, respectively. Finally, the projected basis wave functions are superposed as
\begin{equation}
\Psi^{J\pi} = \sum_{i,K} g_{iK} \Phi_{MK}^{J\pi}(\beta_i)~,
\end{equation}
where the coefficients $g_{iK\alpha}$ are obtained by solving the Hill-Wheeler equation~\cite{HillWheeler1953}.

To quantitatively evaluate the cluster formation probability in the nucleus, we calculate the RWA~\cite{Zhao2021Alpha}. It is defined by the overlap between the wave functions of the mother nucleus and the residues, which represents the probability amplitude of finding a cluster in the mother nucleus. In the case of a cluster with $C_1$ nucleons, the RWA is defined as
\begin{equation}
\begin{aligned}
a&\mathcal{Y}_{L}(a) = a\sqrt{\frac{A!}{(1+\delta_{C_1C_2})C_1!C_2!}}\times \\
&\quad \left\langle \frac{\delta(r-a)}{r^2}\left[Y_{L}(\hat{r})\otimes\left[\Phi^{j_1\pi_1}_{C_1}\otimes\Phi^{j_2\pi_2}_{C_2}\right]_{j_{12}}\right]_{JM} \middle| \Phi^{J\pi}_{M} \right\rangle~,
\end{aligned}
\end{equation}
where $a$ is the distance between the cluster and the residue. $\Phi^{j_1\pi_1}_{C_1}$, $\Phi^{j_2\pi_2}_{C_2}$, and $\Phi^{J\pi}_{M}$ are the total wave functions of the cluster, the residue, and the mother nucleus, respectively. $L$ is the relative angular-momentum between the cluster and the residue, which is coupled with the angular-momentum $j_{12}$ to the total angular-momentum $J$ of the nucleus. There might be several available channels with different relative angular-momentum $L$ for a certain initial state and residue state. We will discuss it in detail later in the next section.

The RWA can be regarded as a wave function of the composed system. Therefore, the radial integral of the squared RWA is a experimental observable, which is called spectroscopic factor (SF)
\begin{equation}
S_L = \int_0^\infty r^2  \mathcal{Y}^2_L(r)dr~.
\end{equation}
The SF provides a quantitative measure of the cluster formation probability, which can be directly measured from the reaction experiments. In this work, we are aiming to investigate the neutron-proton asymmetry on the $^3$H clustering in Boron isotopes. In this sense, we will calculate the SF for $^3$H cluster. The SF of $\alpha$ cluster will also be calculated to make the comparison.

\section{Results and Discussions}
\label{sec:results}
\subsection{Wave Function}

This section begins with the chosen Boron isotopes for the present study. The Boron isotopes, containing $Z = 5$ protons, provide a suitable and relatively simple system to investigate $\alpha$ or $^{3}$H clustering. From the perspective of the cluster model, it is assumed that the five protons in these isotopes can be arranged into two $\alpha$ clusters and one $^{3}$H cluster. This work investigates how the increasing neutron number affects the formation of the $\alpha$ and $^{3}$H clusters, starting with the $\alpha+\alpha+{}^3$H system, i.e., $^{11}$B. The isotopes up to $^{14}$B are assumed to share a similar cluster structure coupled with a varying number of valence neutrons. However, the case of $^{15}$B is distinct. The daughter nuclei corresponding to $\alpha$ or $^{3}$H emission from $^{15}$B are $^{11}$Li and $^{12}$Be, respectively. The ground states of these two nuclei are well known for exhibiting the breaking of the $N=8$ magic number, where the valence neutrons predominantly occupy the $sd$-shell instead of completing the $p$-shell closure. As this shell-structure phenomenon is beyond the scope of our investigation, our analysis is confined to the isotopes from $^{11}$B to $^{14}$B.

To investigate cluster formation in the Boron isotopes, we calculate the RWAs for the $\alpha$ and $^3$H emission channels. This calculation requires the wave functions of both the parent nuclei, $^{11-14}$B, and the corresponding daughter nuclei, $^{7-10}$Li and $^{8-11}$Be. All wave functions are computed within the AMD framework. To validate our calculations, the resulting ground-state energies are compared with experimental data in Table~\ref{tab:energy}, which also lists the corresponding reaction Q-values.
\begin{table*}[t!]  
\centering
\begin{ruledtabular}
\begin{tabular}{ccccc}
 &$^{11}\text{B}(3/2^-)$ &$^{12}\text{B}(1^+)$ &$^{13}\text{B}(3/2^-)$ &$^{14}\text{B}(2^-)$ \\
\hline
Exp. &$-76.21$ &$-79.57$ &$-84.45$ &$-85.42$ \\
Cal. &$-78.56$ &$-77.59$ &$-86.08$ &$-86.67$ \\
\hline\hline
$^{3}${H} channel &$^{8}\text{Be}(0^+)$ &$^{9}\text{Be}(3/2^-)$ &$^{10}\text{Be}(0^+)$ &$^{11}\text{Be}(1/2^+)$ \\
\hline
Exp.& $-56.50(11.23)$ & $-58.16(12.93)$ & $-64.98(10.99)$ & $-65.48(11.46)$ \\
Cal.& $-57.57(13.91)$ & $-55.74(14.77)$ & $-66.48(12.52)$ & $-66.09(13.50)$ \\
\hline\hline
$\alpha$ channel &$^{7}\text{Li}(3/2^-)$ &$^{8}\text{Li}(2^+)$ &$^{9}\text{Li}(3/2^-)$ &$^{10}\text{Li}(2^-)$ \\
\hline
Exp.& $-39.25(8.66)$ & $-41.28(9.99)$ & $-45.34(10.81)$ & $-45.31(11.81)$ \\
Cal.& $-40.53(8.35)$ & $-43.40(4.67)$ & $-46.34(10.06)$ & $-44.38(12.61)$ \\
\end{tabular}
\end{ruledtabular}
\caption{Experimental (Exp) and calculated (Cal) ground state energies for the Boron, Beryllium, and Lithium isotopes relevant to the $^{3}$H and $\alpha$ RWA calculations. All units are in MeV. Q-values for the respective decay channels are given in parentheses.}
\label{tab:energy}
\end{table*}
The calculated ground-state energies exhibit good agreement with the experimental values. This agreement is particularly noteworthy since a single, consistent Hamiltonian is employed for all nuclei under investigation. Such consistency affirms the reliability of the AMD framework for this mass region and, more importantly, validates the obtained wave functions for the subsequent analysis of RWA and SF.

\subsection{RWA and SF}

We calculate the RWA to systematically analyze the possibility of forming a cluster in the parent nuclei. In the calculation of RWA, we need to consider the channels with different relative angular-momentum $L$. The available values of $L$ are determined by the angular-momenta of the parent nucleus and the residue nucleus, as well as the parities of them. In short, $L$ can be $|J_m-J_{r}-J_{c}|\le L\le|J_m+J_{r}+J_{c}|$, where $J_m$, $J_r$, and $J_c$ are the angular-momenta of the parent nucleus, residue nucleus, and the cluster, respectively. Meanwhile, the conservation of parity determines if $L$ is even or odd. For example, for the RWA of $^3$H for $^{14}$B ($^{14}\text{B}(2^-)\to {}^{11}\text{Be}(1/2^+)+^3\text{H}(1/2^+)$), the value of relative angular momentum $L$ is limited in $|2-1/2-1/2|\le L\le|2+1/2+1/2|$ but must be odd because of the parities of all the states. Therefore, $L$ can be $1$ or $3$ in this case.

\begin{figure*}[t!]
\centering
\includegraphics[width=1.0\textwidth]{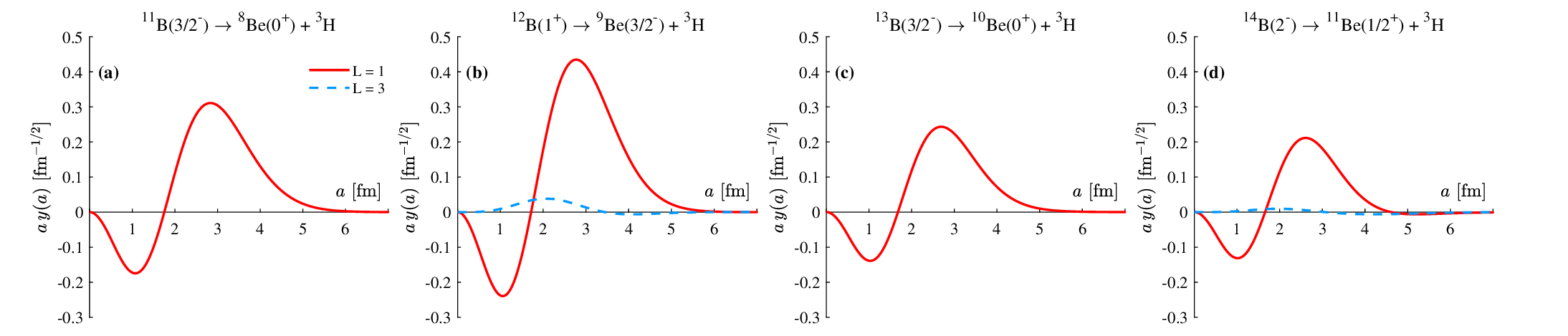}
\includegraphics[width=1.0\textwidth]{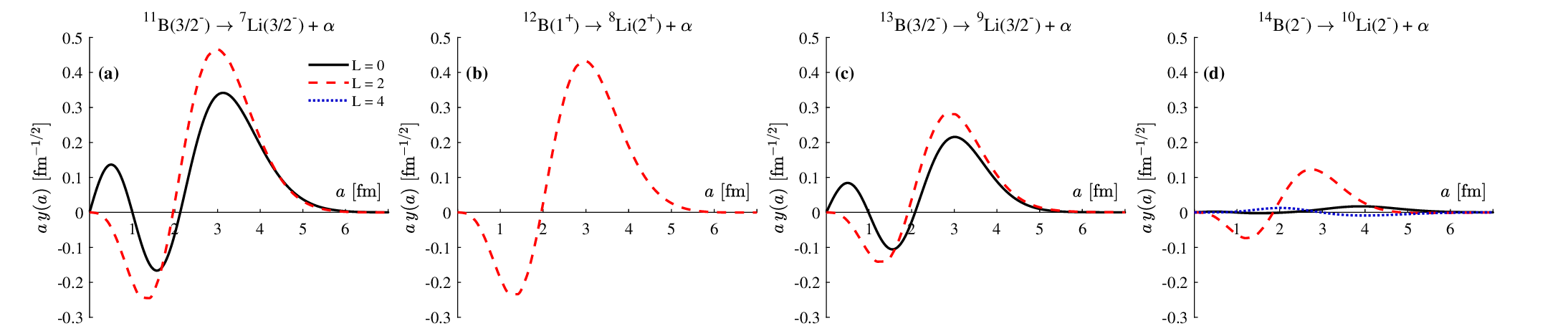} 
\caption{RWAs of $^{3}$H (upper panels) and $\alpha$ (lower panels) in Boron isotopes. The types of curves denote the different relative angular-momentum channels.}
\label{fig:rwa}
\end{figure*}

The calculated RWAs for the $^{3}$H and $\alpha$ clusters in Boron isotopes are displayed in Fig.~\ref{fig:rwa}.
A qualitative inspection reveals that the RWA amplitudes for $^{3}$H are of a comparable magnitude to those for the $\alpha$ cluster. Furthermore, a prominent feature is observed for both types of clusters: their RWAs consistently peak near the nuclear surface, at a radius of approximately 3~fm. This characteristic indicates that the formation of the $^{3}$H cluster, similar to that of the $\alpha$ cluster, is predominantly a surface phenomenon in these nuclei.

\begin{figure}[t!]
    \centering
    \includegraphics[width=0.45\textwidth]{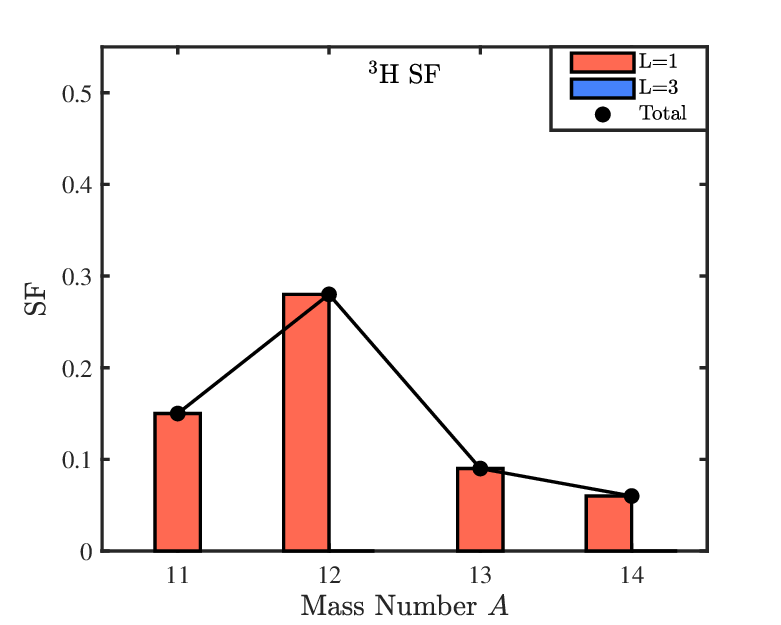}  
    \caption{$^3$H SF for Boron isotopes (mass number \(A = 11\)–\(14\)). The bar charts show the results for different relative angular momentum channels. The dots represent their sum.}
    \label{fig:3h_sf}
\end{figure}

\begin{figure}[t!]
    \centering
    \includegraphics[width=0.45\textwidth]{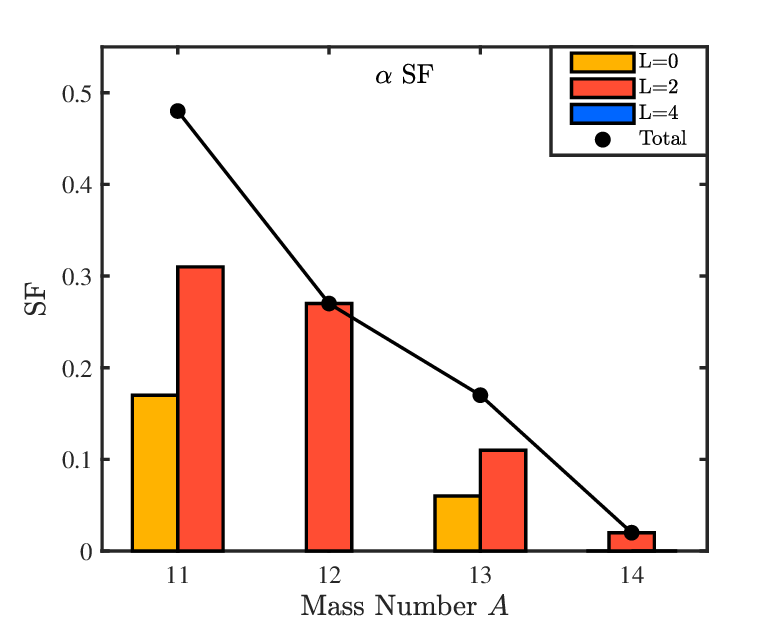}  
    \caption{$\alpha$ SF for Boron isotopes (mass number \(A = 11\)–\(14\)). The bar charts show the results for different relative angular momentum channels. The dots represent their sum.}
    \label{fig:a_sf}
\end{figure}

The SF, which quantifies the cluster pre-formation probability, is obtained by integrating the squared RWA. The calculated SFs for the Boron isotopes from $^{11}$B to $^{14}$B are presented in Fig.~\ref{fig:3h_sf} for the $^3$H cluster and Fig.~\ref{fig:a_sf} for the $\alpha$ cluster. The contributions from different relative angular momenta are shown as bars, with their sum indicated by dots.
These figures reveal contrasting trends in the formation probabilities of the $^3$H and $\alpha$ clusters as the neutron number increases. The SF for the $\alpha$ cluster monotonically decreases with increasing neutron number. This behavior is consistent with the established understanding that a thicker neutron skin suppresses the formation of symmetric clusters like $\alpha$~\cite{Tanaka2021Formation}. In contrast, the SF for the $^3$H cluster exhibits a non-monotonic trend, initially increasing from $^{11}$B to $^{12}$B before decreasing for heavier isotopes. While the suppressive effect of the neutron skin is also expected to influence the $^3$H cluster, the initial increase in its formation probability points to a competing mechanism. This suggests that, in addition to the suppression from the neutron skin, the formation of the neutron-rich $^3$H cluster is concurrently enhanced by the growing neutron-proton asymmetry of the parent nucleus.

\subsection{Enhancement effect by the neutron-proton asymmetry}

To disentangle the competing effects of neutron-skin suppression and isospin-asymmetry enhancement discussed in the previous section, we propose a phenomenological factorization of the SF:
\begin{equation}
S(N,Z,\sigma_c) = N_{\text{skin}}(N,Z) \times F_{\text{iso}}(N,Z,\sigma_c)~,
\end{equation}
where $N_{\text{skin}}(N,Z)$ is a suppression factor that accounts for the influence of the neutron skin, dependent only on the parent nucleus composition $(N,Z)$. The term $F_{\text{iso}}(N,Z,\sigma_c)$ is an enhancement factor that reflects the favorability of forming a cluster with a specific neutron-proton asymmetry, $\sigma_c = (n_c - z_c)/A_c$, within the parent nucleus.

\begin{figure}[t!]
    \centering
    \includegraphics[width=0.45\textwidth]{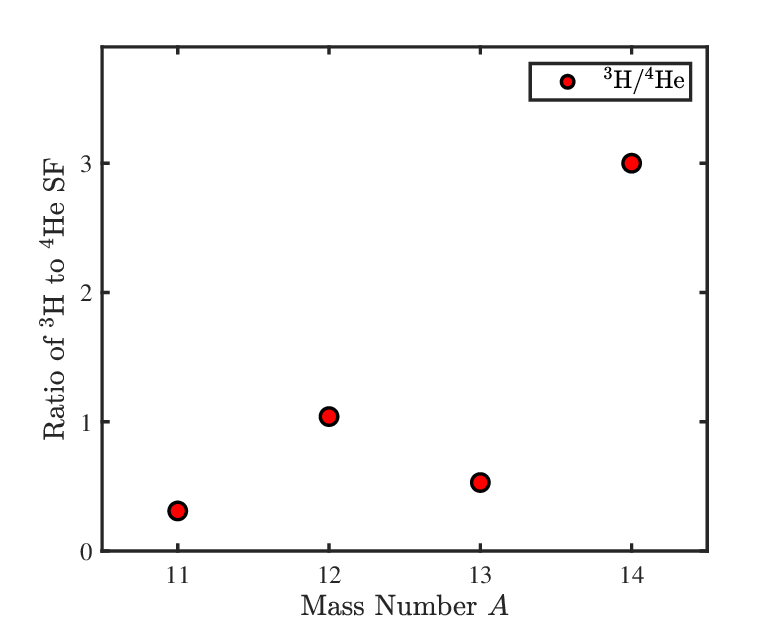}  
    \caption{Ratio of the SF for $^3$H to that for $^4$He as a function of the mass number $A$ of the Boron isotopes.}
    \label{fig:ratio}
\end{figure}

Within this framework, the ratio of the SFs for the $^3$H and $\alpha$ clusters in the same parent nucleus effectively cancels the common suppression factor $N_{\text{skin}}$, thereby isolating the enhancement effect pertinent to the asymmetric $^3$H clustering. This ratio is plotted as a function of the mass number in Fig.~\ref{fig:ratio}.
The plot reveals a generally increasing trend, indicating that the relative formation probability of $^3$H compared to $\alpha$ grows with the neutron number. A notable exception is the dip at $^{13}$B, which can likely be attributed to the shell closure at $N=8$ in its daughter nucleus $^{10}$Be. Overall, these results strongly support our central hypothesis: the formation of the neutron-rich $^3$H cluster is enhanced by the increasing neutron-proton asymmetry of the surrounding nuclear medium. The numerical values used to generate Fig.~\ref{fig:ratio} are detailed in Table~\ref{tab:num_sf}.
\begin{table}[t!]
\centering
\small  
\begin{ruledtabular}
\renewcommand{\arraystretch}{1.2}
\begin{tabular}{lcccc}
 & \textbf{$^{11}$B} & \textbf{$^{12}$B} & \textbf{$^{13}$B} & \textbf{$^{14}$B} \\
\hline
\textbf{L=1}   & $0.15$ & $0.28$        & $0.09$ & $0.06$ \\
\textbf{L=3}   & /    & $\approx 0$ & /    & $\approx 0$ \\
\textbf{Total1}& $0.15$ & $0.28$        & $0.09$ & $0.06$ \\
\hline
\textbf{L=0}   & $0.17$ & /           & $0.06$ & $\approx 0$ \\
\textbf{L=2}   & $0.31$ & $0.27$        & $0.11$ & $0.02$  \\
\textbf{L=4}   & /    & /           & /    & $\approx 0$ \\
\textbf{Total2}& $0.48$ & $0.27$        & $0.17$ & $0.02$ \\
\hline
\textbf{Total1/Total2} & $0.31$ & $1.04$& $0.53$ & $3.00$ \\
\end{tabular}
\end{ruledtabular}
\caption{The numerical results of the SFs shown in Fig.~\ref{fig:3h_sf} and Fig.~\ref{fig:a_sf}.}
\label{tab:num_sf}
\end{table}

This analysis also offers a valuable suggestion for experimental investigations. While absolute SF values are direct observables, their extraction from reaction data can be subject to significant systematic and model-dependent uncertainties. In contrast, measuring the ratio of cross-sections, which corresponds to the SF ratio for an exotic cluster and a well-understood cluster like $\alpha$, can mitigate many of these common uncertainties. Therefore, this ratio serves as a more robust observable for probing the unique structural properties driven by neutron-proton asymmetry.

\section{Summary}
\label{sec:summary}
In this work, we systematically investigate the influence of neutron-proton asymmetry on asymmetric cluster formation by comparing $^3$H and $\alpha$ clustering in the Boron isotopic chain ($^{11-14}$B) using the AMD framework. Our calculations of the SF, which quantify the cluster formation probabilities, reveal two distinct trends. The $\alpha$ cluster SF monotonically decreases with increasing neutron number, a behavior consistent with the established suppression effect from the neutron skin. In stark contrast, the $^3$H cluster SF exhibits a non-monotonic behavior, peaking at $^{12}$B.

This opposing behavior provides clear evidence for a competition between two mechanisms governing the formation of the $^3$H cluster: the suppression caused by the growing neutron skin and an enhancement driven by the increasing neutron-proton asymmetry of the parent nucleus. To isolate and quantify this enhancement, we analyzed the ratio of the spectroscopic factors, SF($^3$H)/SF($\alpha$). This procedure effectively cancels the common neutron-skin suppression effect. The ratio exhibits a generally increasing trend with neutron number, thus demonstrating that a greater neutron-proton asymmetry in the parent nucleus favors the formation of the asymmetric $^3$H cluster.

Finally, we propose that this SF ratio serves as a robust experimental observable. Being less susceptible to systematic uncertainties than absolute SF measurements, it offers a powerful tool for verifying the predicted interplay of competing effects in exotic nuclear clustering.

\begin{acknowledgments}
This work was supported by National Natural Science Foundation of China [Grant Nos. 12465019, 12305123 and 12405147], by the National Key Research and Development Program of China (Grant Nos. 2023YFA1606402 and 2025YFA1614300), by the Guangxi Natural Science Foundation (No. 2023GXNSFDA026005). Numerical calculations were performed in the Cluster-Computing Center of School of Science (C3S2) at Huzhou Normal University.
\end{acknowledgments}

\nocite{*}

\clearpage 
\bibliography{apssamp}

@article{Ikeda1968,
  author = {K. Ikeda and N. Takigawa and H. Horiuchi},
  journal = {Prog. Theor. Phys. Suppl.},
  pages = {464},
  title = {The Systematic Study of the Cluster Structure of Light Nuclei},
  volume = {68},  
  year = {1968}
}

@article{Freer2018,
  author = {M. Freer and H. Horiuchi and Y. Kanada-En'yo and D. Lee and U.-G. Mei{\ss}ner},
  journal = {Rev. Mod. Phys.},
  pages = {035004},
  title = {Microscopic clustering in light nuclei},
  volume = {90},
  year = {2018}
}

@article{Horiuchi1991,
  author = {H. Horiuchi},
  journal = {Nucl. Phys. A},
  pages = {257},
  title = {Microscopic study of clustering phenomena in nuclei}, 
  volume = {522},
  year = {1991}
}

@article{Oertzen2006,
  author = {W. von Oertzen and M. Freer and Y. Kanada-En'yo},
  journal = {Phys. Rep.},
  pages = {43},
  title = {Nuclear clusters and molecules}, 
  volume = {432},
  year = {2006}
}

@article{Tanihata1992Revelation,
  author   = {I. Tanihata and D. Hirata and T. Kobayashi and S. Shimoura and K. Sugimoto and H. Toki},
  title    = {Revelation of thick neutron skins in nuclei},
  journal  = {Phys. Lett. B},
  volume   = {289},
  number   = {3--4},
  pages    = {261--266},
  year     = {1992},
  month    = sep,
  doi      = {10.1016/0370-2693(92)91216-V}
}

@article{Brown2000Neutron,
  author   = {B. Alex Brown},
  title    = {Neutron Radii in Nuclei and the Neutron Equation of State},
  journal  = {Phys. Rev. Lett.},
  volume   = {85},
  number   = {25},
  pages    = {5296--5299},
  year     = {2000},
  month    = dec,
  doi      = {10.1103/PhysRevLett.85.5296}
}

@article{Yoshida2018Investigation,
  author   = {Kazuki Yoshida and Kazuyuki Ogata and Yoshiko Kanada-En'yo},
  title    = {Investigation of \ensuremath{\alpha} clustering with knockout reactions},
  journal  = {Phys. Rev. C},
  volume   = {98},
  number   = {2},
  pages    = {024614},
  year     = {2018},
  month    = aug,
  doi      = {10.1103/PhysRevC.98.024614}
}

@article{Tanaka2021Formation,
  author   = {Junki Tanaka and Zaihong Yang and Stefan Typel and others},
  title    = {Formation of \ensuremath{\alpha} clusters in dilute neutron-rich matter},
  journal  = {Science},
  volume   = {371},
  number   = {6526},
  pages    = {260--264},
  year     = {2021},
  month    = jan,
  doi      = {10.1126/science.abe4688}
}

@article{Kimura2016Antisymmetrized,
  author   = {M. Kimura and T. Suhara and Y. Kanada-En'yo},
  title    = {Antisymmetrized molecular dynamics studies for exotic clustering phenomena in neutron-rich nuclei},
  journal  = {Eur. Phys. J. A},
  volume   = {52},
  number   = {12},
  pages    = {373},
  year     = {2016},
  month    = dec,
  doi      = {10.1140/epja/i2016-16373-9}
}

@article{Chiba2017Laplace,
  author   = {Yohei Chiba and Masaaki Kimura},
  title    = {Laplace expansion method for the calculation of the reduced-width amplitudes},
  journal  = {Prog. Theor. Exp. Phys.},
  volume   = {2017},
  number   = {5},
  pages    = {053D01},
  year     = {2017},
  month    = may,
  doi      = {10.1093/ptep/ptx063},
}

@article{Zhao2021Alpha,
  author   = {Q. Zhao and Y. Suzuki and J. He and B. Zhou and M. Kimura},
  title    = {$\alpha$ clustering and neutron-skin thickness of carbon isotopes},
  journal  = {Eur. Phys. J. A},
  volume   = {57},
  number   = {5},
  pages    = {157},
  year     = {2021},
  month    = may,
  doi      = {10.1140/epja/s10050-021-00465-0},
}

@article{Decharge1980Hartree,
  author   = {J. Decharg{\'e} and D. Gogny},
  title    = {Hartree-{F}ock-{B}ogolyubov calculations with the {D1} effective interaction on spherical nuclei},
  journal  = {Phys. Rev. C},
  volume   = {21},
  number   = {4},
  pages    = {1568--1593},
  year     = {1980},
  month    = apr,
  doi      = {10.1103/PhysRevC.21.1568}
}

@article{Kimura2004,
  author   = {M. Kimura},
  title    = {Deformed-basis antisymmetrized molecular dynamics and its application to $^{20}\mathrm{Ne}$},
  journal  = {Phys. Rev. C},
  volume   = {69},
  number   = {4},
  pages    = {044319},
  year     = {2004},
  month    = apr,
  doi      = {10.1103/PhysRevC.69.044319},
  publisher = {American Physical Society}
}

@article{KanadaEnyo2003,
  author = {Y. Kanada-En'yo and H. Horiuchi},
  journal = {Phys. Rev. C},
  pages = {014319},
  title = {Cluster structures of $\text{Be}$ isotopes},
  volume = {68},
  year = {2003}
}

@article{Tanihata1985,
  author = {I. Tanihata and others},
  journal = {Phys. Rev. Lett.},
  pages = {2676},
  title = {Measurements of Interaction Cross Sections and Nuclear Radii in the Light $p$-Shell Region}, 
  volume = {55},
  year = {1985}
}

@article{KanadaEnyo1995,
  author = {Y. Kanada-En'yo and H. Horiuchi and A. Ono},
  journal = {Phys. Rev. C},
  pages = {628},
  title = {Structure of $\text{Li}$ and $\text{Be}$ isotopes studied with antisymmetrized molecular dynamics}, 
  volume = {52},
  year = {1995}
}

@article{Suhara2010,
  author = {T. Suhara and Y. Kanada-En'yo},
  journal = {Phys. Rev. C},
  pages = {044301},
  title = {Cluster structure of $^{14}\text{C}$ and $^{16}\text{C}$},
  volume = {82},
  year = {2010}
}

@article{HillWheeler1953,
  author = {D. L. Hill and J. A. Wheeler},
  journal = {Phys. Rev.},
  pages = {1102},
  title = {Nuclear Constitution and the Interpretation of Fission Phenomena},
  volume = {89},
  year = {1953}
}

@article{KanadaEnyo2012,
  author = {Y. Kanada-En'yo and M. Kimura and A. Ono},
  journal = {Prog. Theor. Exp. Phys.},
  pages = {01A202},
  title = {Antisymmetrized molecular dynamics and its applications to cluster phenomena},
  volume = {2012},
  year = {2012}
}

\end{document}